\documentclass[12pt]{article}
\usepackage[T1]{fontenc}
\usepackage{stix}
\usepackage{helvet}
\usepackage[utf8]{inputenc}
\usepackage[english]{babel}
\usepackage{amsmath}
\usepackage{amssymb}
\usepackage{mathrsfs}
\usepackage{xcolor}
\usepackage{float}
\usepackage{graphicx}
\usepackage{url}
\usepackage{hyperref}
\hypersetup{hidelinks}

\usepackage[all]{hypcap}
\usepackage{sbc-template}
\usepackage{enumitem}
\usepackage{wrapfig}
\usepackage{multicol}
\usepackage{verbatim}
\usepackage{mdframed}
\usepackage[frozencache=true,cachedir=_minted]{minted2}
\definecolor{bg}{rgb}{0.95,0.95,0.95}
\definecolor{rl}{rgb}{0.5,0.5,0.5}
\newmintinline[umllm]{text}{escapeinside=!!}
\newmintedfile[umllmfile]{sh}{%
  bgcolor=bg,
  breakanywhere=true,
  breaklines,
  fontsize=\footnotesize,
  linenos=false,
  numbersep=3pt,
  python3=true,
  tabsize=2,
  escapeinside=!!}
\usepackage{tikz}
\usetikzlibrary{
  arrows,
  arrows.meta,
  calc,
  chains,
  fit,
  math,
  matrix,
  positioning,
  shapes.geometric,
}
\pgfdeclarelayer{background}
\pgfdeclarelayer{foreground}
\pgfsetlayers{background,main,foreground}
\tikzset{
  x=1em,
  y=1em,
  >=stealth',
  node distance=2em,
}
\def\Q#1{\ensuremath{\mathtt{Q}_{\mathtt{#1}}}}
\def\S#1{\ensuremath{\mathtt{S}_{\mathtt{#1}}}}
\def\B{\ensuremath{\text{\textvisiblespace}}}

\def\R{\ensuremath{\mathtt{R}}}
\newcommand*{\UMLLM}{\url{https://github.com/PUC-Behring-AI/umllm}}

\newcommand*{\GEMINI}{Gemini~2.5 Pro}

\newcommand*{\GPTiv}{GPT~5.4}
\begin{document}
\title{An Experiment on the Simulation of Deterministic Turing Machines by Large Language Models\thanks{%
    This work was published in the Proceedings of the 23rd National Meeting on Artificial and Computational Intelligence, Cuiabá, Brazil.
    \url{doi:10.5753/eniac.2026.30248}}}
\author{%
  Guilherme Lima\inst{1,2},
  Cleyton Slaviero\inst{3},
  Edward H. Haeusler\inst{2}}
\address{%
  PUC-Behring Institute for AI -- PUC-Rio\\
  Rio de Janeiro, Brazil
  \nextinstitute
  Department of Informatics -- PUC-Rio\\
  Rio de Janeiro, Brazil
  \nextinstitute
  Faculdade de Ciência e Tecnologia\\
  Universidade Federal de Mato Grosso -- Campus Várzea Grande\\
  Cuiabá, Brazil
  \email{glima@puc-rio.br, cleyton.slaviero@ufmt.br, hermann@inf.puc-rio.br}}
\maketitle
\begin{abstract}
  Recent studies have shown that the use of chain-of-thought (CoT) under suitable assumptions enables large language models (LLMs) and transformers in general to achieve Turing completeness.
  In this paper, we put this claim to the test.
  We evaluate empirically the feasibility of using state-of-art LLMs with zero-shot CoT prompting to simulate arbitrary deterministic Turing machines.
  We tested four flagship models (DeepSeek V4 Pro, Gemini 2.5 Pro, GPT~5.4 and~5.2) over 116 randomly generated Turing machines with varying configurations.
  As expected, accuracy drops as the number of transitions of the simulated machine grows, but some models/modes of operation perform notably better than others, which corroborates the results of a related benchmark.
\end{abstract}
\section{Introduction}%
\label{sec:introduction}

Large language models (LLMs) have been shown to exhibit remarkable performance in complex reasoning tasks, often far surpassing those of traditional machine learning.
The use of chain-of-thought (CoT), a technique in which the model is instructed to generate a series of intermediate steps, has been demonstrated to greatly improve LLM's reasoning abilities.
In particular, CoT make it possible for LLMs and transformers in general to simulate complex computational tasks and, under suitable assumptions, achieve Turing completeness~\cite{Perez-J-2021,Li-Z-2024,Merrill-W-2024,Jiang-H-2026}.
These results have motivated a growing body of research on the algorithmic capabilities and limitations of LLMs~\cite{LaMalfa-E-2024,Zhou-H-2024,Hua-Z-2026}.

While LLMs may be theoretically capable of simulating arbitrary computations, their execution (inference) is inherently stochastic and therefore prone to errors.
Practical limitations such as the maximum size of context windows can also hinder the feasibility of such simulations, even in a CoT setting.
Other important constraints are the time it takes for the LLM to actually perform the simulation and the monetary cost of doing so, especially considering the steep prices of flagship models.

In this paper, we investigate the feasibility of simulating arbitrary computational tasks using off-the-shelf LLMs with zero-shot chain-of-thought prompting.
By ``off-the-shelf LLMs'' we mean recent models from popular providers and by ``zero-shot'' we mean without using examples.
The ultimate test of computational ability is \emph{Turing completeness}, or the capacity of simulating an arbitrary Turing machine.
In a sense, what we are evaluating in this paper is the extent to which using a natural language prompt as the executable specification of a computational task is feasible in practice.
The ``computational task'' in this case is running a deterministic Turing machine over a given input with the LLM playing the role of the universal machine.

To evaluate the ability of LLMs to simulate deterministic Turing machines, we proceed as follows.
First, we use a system prompt to establish the format to be used to represent Turing machines and the steps required to compute a transition.
This is done in an abstract manner, without referring to any specific machine.
Then, in a human prompt, we present a specific machine to be simulated together with its input data.
From this point on, we enter a loop in which we ask the LLM to execute one transition of the simulated machine and check the model's output at each iteration.
Eventually, either the LLM produces an invalid transition or it reaches a state in which the simulated machine is supposed to halt.
In either case, we stop the evaluation and record the total number of correct transitions.

We used the previous approach to evaluate four recent LLMs (DeepSeek's~V4 Pro, Google's \GEMINI, OpenAI's \GPTiv\ and~5.2) over 116 randomly generated Turing machines with varying configurations.
These machines and the Python framework used to simulate their execution via LLMs, called UMLLM, are made available under an open-source license (Apache 2.0).\footnote{\UMLLM}
\GEMINI\ performed notably better than the other models, which corroborates the results observed in a related benchmark~\cite{Wu-H-2025}.
As expected, performance deteriorates significantly as the number of transitions grows.
For example, from 60 transitions onward, the accuracy of \GEMINI\ drops below 25\%, i.e., it predicts a correct transition only about 25\% of the time.

The rest of the paper is organized as follows.
Section~\ref{sec:background} presents background on Turing machines and discusses related work.
Section~\ref{sec:implementation} presents our approach to simulate the execution of Turing machines using LLMs with zero-shot CoT prompting.
Section~\ref{sec:evaluation} presents the experimental evaluation, including the dataset, metrics, setup, and results.
Section~\ref{sec:conclusion} presents our conclusions and discusses some future work.




\section{Background}%
\label{sec:background}

The \emph{Turing machine} is a mathematical model of computation introduced by Alan M.~Turing~\cite{Turing-A-M-1936}.
Formally, a Turing machine is a quadruple $M=(K,\Sigma,\delta,s,h)$ where~$K$ is a finite set of states, $\Sigma$ is a finite set of symbols, $\delta$ is the transition function, and $s$ and $h$ are the initial and halting states.\footnote{This is a slight variation of the definition found in~\cite{Papadimitriou-C-H-1994}.}
Intuitively, the machine consists of a current state, an infinite one dimensional tape of symbols, and a read-write head.
Function $\delta$ is the ``program'' of the machine.
It specifies, for each combination of current state $q\in{K}$ and current symbol $\sigma\in\Sigma$, a triple $\delta(q,\sigma)=(p,\rho,D)$, where $p$ is the next state, $\rho$ is the symbol to be overwritten on $\sigma$, and $D$ is either ``$L$'' (left) or ``$R$'' (right).

A Turing machine $M$ starts in the initial state $s$ with some finite input string of symbols written on its tape and the head at some initial position.
It then makes a transition according to $\delta$, changing its state, writing a symbol, and moving its head left or right.
This process repeats until the machine reaches state $h$, at which point it halts.
The content of the tape at this moment is the machine's output.
If the machine halts on a given input $x$, we can define the output of the machine $M$ on $x$, denoted $M(x)$.
It is possible that $M$ never halts on input $x$, in which case its output for $x$ undefined, i.e., $M(x)=\mathop{\uparrow}$.

The description above assumes that $\delta$ is a total function, which means that at each step during the computation the machine has a single uniquely defined next step.
This type of machine is called \emph{deterministic}, as opposed to non-deterministic ones in which there may be more than one appropriate next step.
In this paper, we deal only with deterministic Turing machines.

One fundamental fact about Turing machines, established by Turing in its seminal paper, is the existence of universal Turing machines.
A \emph{universal Turing machine} $U$ is a Turing machine that interprets its input as the description of another Turing machine $M$ concatenated with the description of an input $x$ to that machine.
$U$'s function is then to simulate the behavior of $M$ when presented with input $x$, i.e., $U(M;x)=M(x)$.
The ability to simulate arbitrary Turing machines is precisely what we are trying to achieve here using LLMs.
But before simulating the execution of a Turing machine using an LLM or any other computational process, we need to agree on a format for describing arbitrary machines and their inputs.
We introduce the particular format we adopt in this paper next.

\subsection{Describing and Simulating Arbitrary Turing Machines}%
\label{sec:background:format}

To encode the description of a Turing machine we use the symbols ``$\mathtt{Q}$'', ``$\mathtt{S}$'', ``$\mathtt{0}$'', ``$\mathtt{1}$'', ``$\mathtt{L}$'', ``$\mathtt{R}$'', and ``$\B$''.
The idea is to use strings of the form $\mathtt{Q0}$, $\mathtt{Q1}$, $\mathtt{Q10}$, $\mathtt{Q11}$, \dots\ to represent the states of the machine and strings of the form $\mathtt{S0}$, $\mathtt{S1}$, $\mathtt{S10}$, $\mathtt{S11}$, \dots\ to represent its (non-blank) symbols.
(For readability, we sometimes use subscripts when writing such strings, e.g., $\Q{1011}$ instead of $\mathtt{Q1011}$ and $\S{101}$ instead of $\mathtt{S101}$.)
The symbols $\mathtt{L}$ and $\mathtt{R}$ are used to specify the direction of movement (left or right) of the head when encoding the description of the transition function of the machine.
Finally, symbol $\B$ represents the blank symbol used, among other things, to delimit the written portion of the machine's tape.

We describe the transition function of the simulated Turing machine by giving its transition table.
For example, here is part of the transition table of a Turing machine \textsc{Invert}, which ``inverts'' all bits to the right of the current head position:
\begin{align}
&\Q{0}\S{0}\Q{0}\S{1}\R\\
&\Q{0}\S{1}\Q{0}\S{0}\R\\
&\Q{0}\B\,\Q{1}\B\,\R
\end{align}
This transition table defines three transitions prescribing that:
\begin{enumerate}
\item When in state $\Q{0}$, reading symbol $\S{0}$, the machine shall stay in state $\Q{0}$, write $\S{1}$ over the current head position, and then move the head to the right.
\item When in state $\Q{0}$ reading $\S{1}$, the machine shall stay in $\Q{0}$, write $\S{0}$, and move right.
\item When in state $\Q{0}$ reading $\B$, the machine shall change to $\Q{1}$, write $\B$, and move right.
\end{enumerate}

To describe the content the tape, the current state, and the position of head of the simulated machine, we use strings of the form
\begin{equation}\label{work-tape}
  \B\;
  s_{i}\ldots s_{j}\;
  \mathtt{Q}_{k}\;
  s_{m}\ldots s_{n}\;
  \B
\end{equation}
where each $s$ is either a blank ($\B$) or some non-blank symbol of the machine.
We call such strings \emph{work tapes} and we require that they be delimited by blanks on the extremities, a requirement that must be observed by the simulation process.

A work tape such as~\eqref{work-tape} above indicates three things:
(i)~the machine's current state, namely, $\mathtt{Q}_k$;
(ii)~the written portions of the machine's tape to the left and right of the head, namely, ``$\B\,s_i\ldots{s_j}$'' and ``$s_{m+1}\ldots{s_n}\,\B$'';
and (iii)~the symbol being currently read by the head, namely, $s_m$.
A transition of the machine operates over its current work tape and produces a new work tape.

Besides giving the transition table and initial work tape, the last thing missing to completely describe a Turing machine to be simulated is specifying its final state.
We do this by indicating the particular $\mathtt{Q}_f$ which is to be used to represent the final state.
The machine is assumed to halt whenever $\mathtt{Q}_f$ occurs in the work tape.

\begin{figure}[h]
  \umllmfile{figs/invert.txt}%
  \caption{The \textsc{Invert} Turing machine in UMLLM format.}%
  \label{fig:invert}
\end{figure}

Figure~\ref{fig:invert} gives a complete specification of Turing machine \textsc{Invert} together with some sample input in the format we just described.
This is exactly the format adopted by UMLLM, the framework for simulating Turing machines via LLMs presented in Section~\ref{sec:implementation}.
We call this format the UMLLM format.
It consists of three blocks separated by empty lines.
First, comes the transition table, which is given as described above with any uninteresting transitions omitted if desired.
(If such transitions are ever reached during simulation, they are deemed invalid.)
The second block contains the halting state and the third block contains the initial work tape.
ASCII whitespace within blocks and lines beginning with ``\umllm/#/'' are ignored by the parser.

To simulate the execution of the machine given in Figure~\ref{fig:invert}, we start with the initial work tape and proceed to update it one transition at a time, following the rules prescribed by the transition table.
The diagram in Figure~\ref{fig:trace} displays the work tapes produced by successive transitions of machine \textsc{Invert} when initialized as in Figure~\ref{fig:invert}.
Each transition is indicated by an arrow labeled with the position of the transition in the transition table (first, second, or third).
After nine transitions the machine halts and the content of the work tape is exactly the content of the initial work tape with each bit inverted (i.e., $\S0$ replaced by $\S1$ and $\S1$ replaced by $\S0$).

\begin{figure}[h]
  \centering
  \newcommand*\X[1]{{\color{purple}#1}}%
  \begin{tikzpicture}[node distance=1em,font=\small]
    \begin{scope}[draw]
      \node(T0){$\B\X{\Q0}\S1\S1\S0\S0\S1\S1\S0\S0\B$};
      \node[right=of T0](T1){$\B\S0\X{\Q0}\S1\S0\S0\S1\S1\S0\S0\B$};
      \node[right=of T1](T2){$\B\S0\S0\X{\Q0}\S0\S0\S1\S1\S0\S0\B$};
      \node[below=of T0](T3){$\B\S0\S0\S1\X{\Q0}\S0\S1\S1\S0\S0\B$};
      \node[right=of T3](T4){$\B\S0\S0\S1\S1\X{\Q0}\S1\S1\S0\S0\B$};
      \node[right=of T4](T5){$\B\S0\S0\S1\S1\S0\X{\Q0}\S1\S0\S0\B$};
      \node[below=of T3](T6){$\B\S0\S0\S1\S1\S0\S0\X{\Q0}\S0\S0\B$};
      \node[right=of T6](T7){$\B\S0\S0\S1\S1\S0\S0\S1\X{\Q0}\S0\B$};
      \node[right=of T7](T8){$\B\S0\S0\S1\S1\S0\S0\S1\S1\X{\Q0}\B$};
      \node[below=of T6](T9){$\B\S0\S0\S1\S1\S0\S0\S1\S1\B\X{\Q1}$\rlap{$\B$}};
      \node[right=of T9]{\emph{(halting state reached)}};
    \end{scope}
    \begin{scope}[font=\scriptsize]
      \draw[->](T0) -- node[above]{(2)} (T1);
      \draw[->](T1) -- node[above]{(2)} (T2);
      \draw[->](T2.east) -- node[above]{(1)} ++(1em,0) -- ++(0,-1em)
      -| ($(T3.west)+(-.75em,0)$) -- (T3.west);
      \draw[->](T3) -- node[above]{(1)} (T4);
      \draw[->](T4) -- node[above]{(2)} (T5);
      \draw[->](T5.east) -- node[above]{(2)} ++(1em,0) -- ++(0,-1em)
      -| ($(T6.west)+(-.75em,0)$) -- (T6.west);
      \draw[->](T6) -- node[above]{(1)} (T7);
      \draw[->](T7) -- node[above]{(1)} (T8);
      \draw[->](T8.east) -- node[above]{(3)} ++(1em,0) -- ++(0,-1em)
      -| ($(T9.west)+(-.75em,0)$) -- (T9.west);
    \end{scope}
  \end{tikzpicture}
  \caption{Execution trace of \textsc{Invert}.}%
  \label{fig:trace}
\end{figure}

In Section~\ref{sec:implementation}, we describe an approach to instruct an LLM to execute a sequence of transitions such as that of Figure~\ref{fig:trace} for a given UMLLM specification.
Before that, however, let's review related work on the simulation of computational processes by LLMs.

\subsection{Related Work}

There are two kinds of work related to this paper.
First, there are the theoretical results on the computation ability of LLMs, or more generally, the transformer architecture.
For example, it has been established that transformers can represent arbitrary computations of any finite-state automaton~\cite{Liu-B-2023} and, under suitable assumptions, even simulate arbitrary Turing machines~\cite{Perez-J-2021,Jiang-H-2026}.
It has also been argued that chain-of-thought (CoT), a technique in which the model auto-repressively generates a sequence of intermediate results before giving the final answer, is essential in enabling transformers to perform the serial computations required for solving complex reasoning tasks~\cite{Li-Z-2024}.

The second kind of work, more closely related to this paper, are those on the experimental evaluation of the computational ability of LLMs.
One such work is~\cite{Wu-H-2025} which proposes TMBench, a benchmark for assessing the capacity of LLMs in simulating programs expressed in a Turing-complete formalism.
Although the benchmark is called TMBench, with the ``TM'' standing for Turing machine, the formalism it adopts is not that of Turing machines.
Instead it uses what the authors call $m$-tag systems which are a version of the classical formalism introduced by E.~Post~\cite{Post-E-L-1943}.
A further difference between~\cite{Wu-H-2025} and our work, beyond the computation formalism used, is the prompting technique.
While they use one-shot non-CoT prompting (one example, no chain-of-thought instruction), we use zero-shot CoT prompting.
The experimental results of \cite{Wu-H-2025} are thus not directly comparable to ours but, as we will see in Section~\ref{sec:evaluation}, the best performing model in both benchmarks was \GEMINI.

Besides TMBench, we could find no other benchmark that uses Turing machines or related formalisms for evaluating the computational capacity of LLMs.
There are, of course, many benchmarks for assessing the ability of LLMs in solving tasks that require some form of computation (cf.~\cite{Wang-P-2026}).





\section{Method}%
\label{sec:implementation}

To simulate the execution of a deterministic Turing machine by an LLM, we use single a \emph{system prompt} describing:
\begin{itemize}
\item The format used to encode the simulated machine (see Section~\ref{sec:background:format}).
\item The data-structures to be used during the simulation.  These consist of ten tapes: the Work tape of the simulated machine, a read-only Machine tape containing the transition table, and auxiliary tapes holding intermediate data.
\item The seven steps required to compute a single transition of the simulated machine.  These steps use the aforementioned tapes to construct and apply a single string substitution operation on the Work tape.
\item The format to be used by the LLM to present the results (i.e., the resulting Work tape) and the commands which it should expect to receive in subsequent interactions (either ``continue'' or ``stop'').
\end{itemize}
The actual system prompt we use is shown in Figure~\ref{fig:system}.
In the last paragraph of the prompt, we ask the LLM to explain what it is doing during the execution of each of the seven steps that compute the transition.
This is the chain-of-thought instruction.
Although we are not interested in these explanations, it has been observed that the mere fact of producing them improves the accuracy of the model in tasks of this kind.

\begin{figure}[h]
  \centering
  \scriptsize
  \begin{minipage}[t]{\textwidth}%
    \begin{mdframed}%
    \begin{multicols}{2}%
      \raggedright
      \sffamily
      \input{figs/system.tex}%
    \end{multicols}%
    \end{mdframed}%
  \end{minipage}%
  \caption{System prompt.}%
  \label{fig:system}
\end{figure}

The system prompt is followed by a \emph{human prompt} (or user prompt) that describes the actual Turing machine to be simulated together with its input.
Figure~\ref{fig:human} shows the human prompt corresponding to the machine \textsc{Invert} described in Section~\ref{sec:background:format}.

\begin{figure}[h]
  \centering
  \scriptsize
  \begin{minipage}[t]{\textwidth}%
    \begin{mdframed}
      \raggedright
      \sffamily
      \input{figs/human.tex}
    \end{mdframed}
  \end{minipage}%
  \caption{Human prompt of \textsc{Invert}.}%
  \label{fig:human}
\end{figure}

Note that neither the system prompt nor the human prompt mention the existence of initial or final states.
This is because the goal here is to use the LLM solely to compute transitions of the simulated machine, i.e., constructing and applying a single substitution to the current work tape, as described in the system prompt.
Every time the LLM replies with a new work tape, we process it to see if it is correct (i.e., if it corresponds to the expected one) and then check to see if it contains the halting state.
If so, we stop the execution.
Otherwise, we ask the LLM to perform another transition.

Figure~\ref{fig:ai} displays the response generated by OpenAI's GPT-5.4 after receiving the system and human prompts shown above.
Note that the model adopts the terminology and notation introduced in the system prompt to explain each step it preformed.
The only part of the response the matters to us, however, is the string between the tags \textless{work}\textgreater\ and \textless/{work}\textgreater, which contains the new work tape, obtained after executing a single transition of the simulated machine.
Incidentally, the work tape shown here (highlighted in red) is correct.
It matches exactly the work tape obtained after the first transition (see Figure~\ref{fig:trace}).

\begin{figure}[h]
  \centering
  \scriptsize
  \begin{minipage}[t]{\textwidth}%
    \begin{mdframed}
    \begin{multicols}{2}%
      \raggedright
      \sffamily
      \input{figs/ai.tex}
    \end{multicols}%
    \end{mdframed}
  \end{minipage}%
  \caption{GPT-5.4's response after receiving the system prompt and the human prompt of \textsc{Invert}.}%
  \label{fig:ai}
\end{figure}

\subsection{Implementation}

We implemented the approach described above as a Python library called UMLLM, which is made available under the Apache 2.0 license.\footnote{\UMLLM}
UMLLM is a framework for the execution of Turing machines.
It comes with a traditional Turing machine executor written in Python which implements the same algorithm described in the system prompt (Figure~\ref{fig:system}).
It also comes with CLI and web interfaces that can be used to execute machines in the UMLLM format interactively.

The interesting part of UMLLM is, of course, its support for LLMs.
It provides LLM-backed executors which can be used run Turing machines using the approach described previously.
To interface with the LLMs, UMLLM uses the LangChain\footnote{\url{https://www.langchain.com/}} library, more specifically, its chat model API which provides a uniform message-based interface for invoking language models from different providers.

One important point about LangChain's chat model API is that the core interface for invoking LLMs is stateless.
That is, one sends a series of messages to the LLM and receives back one or more messages comprising the model's reply with no data carrying over to the next invocation, at least not implicitly.
This means that to implement the ``continue'' command described in the system prompt, we have to send to the LLM the full message history.
This history contains the initial system and human prompts, the 1st model reply, the 1st ``continue'' command (which is another human prompt), the 2nd model reply, the 2nd ``continue'' command, and so on.
That said, UMLLM can be configured to \emph{trim} the history to a desired maximum length.
For example, if we set this length to 0, at each LLM invocation, instead of sending the full history plus a ``continue'' command, UMLLM will send only the system and user prompts with the initial work tape updated to reflect the current configuration of the simulated machine.
We compare both modes of operation, full history (untrimmed) versus no history (trimmed), in the evaluation.




\section{Evaluation}%
\label{sec:evaluation}

\newcommand*\pQ{\ensuremath{\mathbf{\#Q}}}
\newcommand*\pW{\ensuremath{\mathbf{\#W}}}
\newcommand*\pT{\ensuremath{\mathbf{\#T}}}
\newcommand*\DSV{\textsf{DSV4}}
\newcommand*\GEM{\textsf{GEM2.5}}
\newcommand*\GPT[1]{\textsf{GPT#1}}
\newcommand*\tr{\textsf{(tr)}}

For the experimental evaluation, we generated a dataset with 116 random Turing machines in the UMLLM format.
The varieties of machines in the dataset are described in Table~\ref{tab:variations}.
Every machine has 2, 4, or 8 states (parameter \pQ) and an alphabet with exactly two non-blank symbols ($\S0$ and $\S1$).
The written portion of their initial work tape contains exactly 8, 16, 32, or 64 symbols (parameter \pW) and each machine is guaranteed to halt after a specific number of transitions, either 8, 16, 32, 64, or 128 (parameter \textbf{\#T}).
The dataset contains two random machines for every combination of these parameters with the exception of $\pQ=2$ and $\pW=8$ with $\pT=64$ or $\pT=128$, for which we could find no examples.
Hence the number $116$ $(=2\times{3}\times{4}\times{5}-4)$.
Figure~\ref{fig:example} illustrates one of the machines in the dataset.

\begin{table}[h]
  \caption{The tested Turing machine variations.}
  \label{tab:variations}
  \centering\footnotesize
  \begin{tabular}{c|l|l}
    & variations & description\\\hline
    \textbf{\pQ} & 2, 4, 8 & number of states\\
    \textbf{\pW} & 8, 16, 32, 64 & length of the initial work tape\\
    \textbf{\pT} & 8, 16, 32, 64, 128 & number of transitions until halting
  \end{tabular}
\end{table}

\begin{figure}[h]
\umllmfile{figs/Q04-S02-W64-C32-aa60516ac1a338df0cdd16f883d0c59491ce44a8afe643fc5e20dc39279a27ef.txt}%
  \caption{One of the Turing machines in the dataset ($\pQ=4$, $\pW=64$, $\pT=32$).}%
  \label{fig:example}
\end{figure}

We tested four different LLMs via the official APIs of their providers: DeepSeek's V4 Pro (\DSV), Google's Gemini 2.5 Pro (\GEM), and OpenAI's GPT 5.4 (\GPT{5.4}) and 5.2 (\GPT{5.2}).
For more information, including prices, see Table~\ref{tab:llms}.
When calling the models, the only API parameters we set explicitly were \emph{temperature} and \emph{seed}, both set to zero.
All other parameters were left with their default values.
(We wanted to test each model in its vanilla configuration, which is the one seen by the average user.)

\begin{table}[h]
  \caption{The tested LLMs (prices in USD, 2026-06-08).}
  \label{tab:llms}
  \centering\footnotesize
  \begin{tabular}{l|l|l|l|r}
    provider
    & model
    & abbrev.
    & snapshot
    & \$/1M in / out tokens\\\hline
    DeepSeek
    & deepseek-v4-pro\footnotemark[5]
    & \DSV
    & 2026-04-24
    & \$0.435 / \$00.87\phantom{0}\\
    Google
    & gemini-2.5-pro\footnotemark[6]
    & \GEM
    & 2025-06
    & \$2.50\phantom{0} / \$15.00\phantom{0}\\
    OpenAI
    & gpt-5.4\footnotemark[7]
    & \GPT{5.4}
    & 2026-03-05
    & \$2.50\phantom{0} / \$15.00\phantom{0}\\
    & gpt-5.2\footnotemark[8]
    & \GPT{5.2}
    & 2025-12-11
    & \$1.75\phantom{0} / \$14.00\phantom{0}
  \end{tabular}
\end{table}
\footnotetext[5]{\url{https://huggingface.co/deepseek-ai/DeepSeek-V4-Pro}}
\footnotetext[6]{\url{https://ai.google.dev/gemini-api/docs/models/gemini-2.5-pro}}
\footnotetext[7]{\url{https://developers.openai.com/api/docs/models/gpt-5.4}}
\footnotetext[8]{\url{https://developers.openai.com/api/docs/models/gpt-5.2}}

We run each model over every machine in the dataset twice.
Once in \emph{full-history mode}, in which on each interaction we would send the full message history, and once in \emph{no-history (trimmed) mode}, in which on each interaction we would send exactly two messages, one with the system prompt and one with the human prompt containing the updated work tape.
This gives a total of $2\times(4\times116)=928$ runs.
The total and average (per run) token consumption and USD expenditure values for each model-mode is shown in Table~\ref{tab:cost}.
We use the suffix \tr\ to distinguish trimmed runs from untrimmed runs.
Note the higher input token consumption figures for untrimmed \GEM\ and \GPT{5.2}.
These indicate that the system was able to maintain longer ``conversations'' (runs) with these models/modes.
We will come back to this point later.

\begin{table}[h]
  \caption{Token consumption and expenditure (USD).}%
  \label{tab:cost}
  \centering\footnotesize
  \begin{tabular}{r|rr|rr|rr|rr}
    &\DSV
    &\tr
    &\GEM
    &\tr
    &\GPT{5.2}
    &\tr
    &\GPT{5.4}
    &(tr)\\\hline
    in tokens
    &5,103K
    &1,448K
    &53,882K
    &2,035K
    &18,120K
    &654K
    &9,448K
    &1,278K\\
    out tokens
    &2,172K
    &5,416K
    &4,265K
    &4,404K
    &670K
    &362K
    &665K
    &565K\\
    \$
    &\$4.11
    &\$5.34
    &\$198.67
    &\$71.14
    &\$41.09
    &\$6.21
    &\$33.59
    &\$11.66\\\hline
    avg. in tokens
    &44K
    &12K
    &465K
    &18K
    &156K
    &6K
    &81K
    &11K\\
    avg. out tokens
    &19K
    &47K
    &37K
    &38K
    &6K
    &3K
    &6K
    &5K\\
    avg. \$
    &\$0.04
    &\$0.05
    &\$1.71
    &\$0.61
    &\$0.35
    &\$0.05
    &\$0.29
    &\$0.10
  \end{tabular}
\end{table}

\subsection{Results}

\newcommand*\Nc{N_{\text{correct}}}
\newcommand*\Nt{N_{\text{total}}}

\begin{figure}[h]
  \centering
  \includegraphics[width=\textwidth,
    clip=true,trim=10 10 10 10]{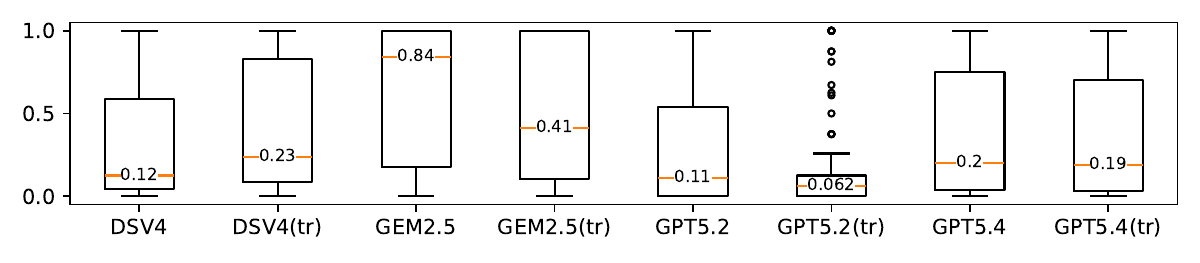}
  \caption{Run progress (for each run, \# correct transitions / \# total transitions).}
  \label{fig:progress}%
\end{figure}

Figure~\ref{fig:progress} shows the progress of all runs grouped by model/mode.
We define the \emph{progress} $P$ of a run $r$ as the number of correct transitions performed by the LLM divided by the total number of transitions required for the machine to halt:
\[
  P(r)\coloneq\frac{\Nc(r)}{\Nt(r)}
\]
Keep in mind that we stop the run as soon as the LLM produces an incorrect transition.
So, $P(r)=0$ means that the LLM failed on the first step, $P(r)=0.5$ means it failed exactly midway during the execution, and $P(r)=1$ means it executed every transition successfully until the end.
As we can see in Figure~\ref{fig:progress}, the values of progress of untrimmed \GEM\ runs are significantly higher than those of the other models.

\begin{figure}[h]
  \centering
  \includegraphics[width=\textwidth,
    clip=true,trim=1 1 1 1]{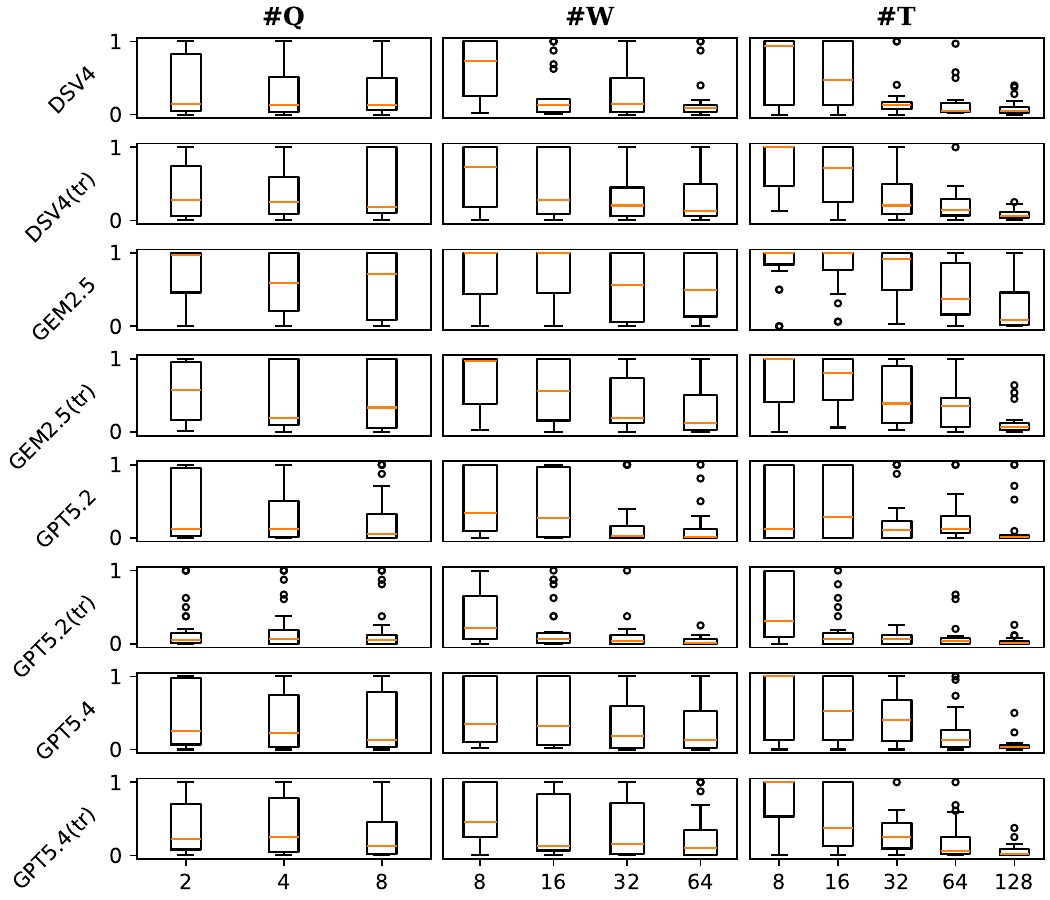}%
  \caption{Run progress (detailed).}%
  \label{fig:progress-detailed}
\end{figure}

A breakdown of run progress across the machine parameters \pQ, \pW, \pT\ is shown in Figure~\ref{fig:progress-detailed}.
One thing we can immediately observe is that, in column \pT, as the number of transitions required for the machine to halt grows, the overall progress values drop significantly across all models.
A similar but less dramatic drop can be observed when the length of the initial work tape (\pW) increases.

Another metric we used to evaluate the LLMs performance was transition-wise accuracy, which quantifies the proportion of correct transitions at a each step of the execution~\cite{Wu-H-2025}.
The \emph{(transition-wise) accuracy} at transition~$i$ of a given set $R$ of runs is defined as
\[
  A_R(i)\coloneq\frac
  {\left|\{r\in{R}\mid\Nc(r)\ge{i}\}\right|}
  {\left|\{r\in{R}\mid\Nt(r)\ge{i}\}\right|}
\]
For example, $A_R(30)=0.5$ means that at step 30 the proportion of correct transitions considering the runs with at least 30 transitions in $R$ is 50\%.

\begin{figure}[h]
  \centering
  \includegraphics[width=\textwidth,
    clip=true,trim=0 0 0 0]{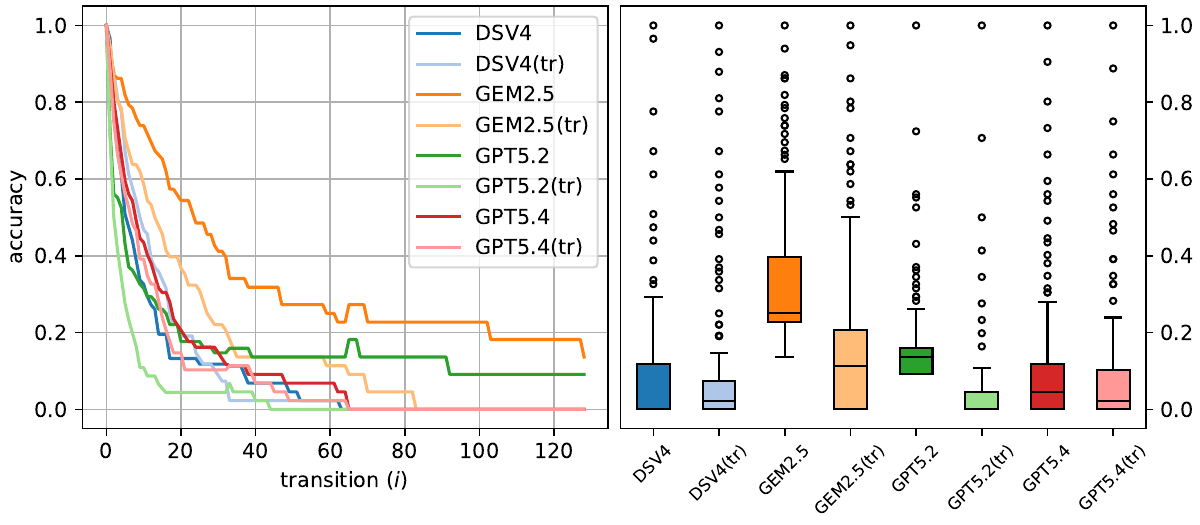}%
  \caption{Transition-wise accuracy.}%
  \label{fig:accuracy}
\end{figure}

Figure~\ref{fig:accuracy} shows the transition-wide accuracy of all runs grouped by model/mode.
The line plots (left) show the accuracy values for each transition~$i$ in the 0--128 range, while the box-plots (right) show aggregated accuracy values for the whole range.
As we can see in the line plots, accuracy drops significantly after~20 transitions, with untrimmed \GEM\ consistently obtaining the best accuracy values\footnotemark\ and, somewhat surprisingly, untrimmed \GPT{5.2} performing better than the remaining models/modes in longer runs.
In fact, only untrimmed \GEM\ and \GPT{5.2} were able to complete runs with~128 transitions, which explains the input token consumption figures of Table~\ref{tab:cost}.

\footnotetext{A similar behavior regarding \GEM\ was observed in the evaluation done in \cite{Wu-H-2025}, which used LLMs for simulating Post systems~\cite{Post-E-L-1943}.}

As we can see in the box-plots, the overall accuracy of untrimmed runs is higher than that of their trimmed counterparts.
Our hypothesis to explain this behavior is the following: the contents of earlier messages present in the history of untrimmed runs, in particular, earlier chain-of-thought excerpts, influence the model and somehow increase its chance of getting the current transition right.


\section{Conclusion}%
\label{sec:conclusion}

In this paper, we presented an empirical evaluation of the simulation of deterministic Turing machines by LLMs.
We proposed an approach for accomplishing such simulations using zero-shot CoT prompting and evaluated it empirically by running four flagship LLMs over 116 randomly generated Turing machines.
The fact that some runs were able to achieve the halting state after 128 transitions (the maximum number of transitions in the dataset) is a testament to the capacity of current LLMs and, to a certain degree, demonstrates the feasibility of our approach.
That said, in general, accuracy drops sharply across all models after~20 transitions and the simulation is by no means efficient or cheap.

In future work, we intend evaluate further models and improve some known shortcomings of our approach.
For example, the current CoT instruction is too simple.
We could try to generate more accurate CoT by being more explicit in its instruction or use examples (few-shot) to increase the chances of eliciting a correct response.
There is also room for improvement in the way history is being handled.
We could use more advanced compression or summarization techniques to reduce the amount of data that is sent to the LLM on each iteration, which would reduce the overall latency and costs.




\bibliographystyle{sbc}
\bibliography{bib}
\end{document}